\documentstyle[amssymb,prb,aps,twocolumn,epsf]{revtex}

\begin{document}
\twocolumn[\hsize\textwidth\columnwidth\hsize\csname@twocolumnfalse\endcsname

\draft

\title{The surface barrier in mesoscopic type I and type II
superconductors}
\author{Alexander D. Hern\'andez$^{1,2}$ and Daniel Dom\'{\i}nguez$^2$}

\address{$^1$Laboratorio de
Superconductividad, Facultad de F\'{\i}sica-IMRE, Universidad de la
Habana, 
10400, Ciudad Habana, Cuba.\\
$^2$Centro At\'{o}mico Bariloche, 8400 San Carlos  de
Bariloche, R\'{\i}o Negro, Argentina}

\maketitle

\begin{abstract} 

We study the surface barrier for magnetic
field penetration in mesoscopic samples of both type I and type II 
superconductors. Our results are obtained from numerical 
simulations of the time-dependent Ginzburg-Landau
equations. We calculate the dependence of the first field for 
flux penetration ($H_p$) with the 
Ginzburg-Landau parameter ($\kappa $) observing an increase of
$H_p$ with decreasing $\kappa$ for a superconductor-insulator
boundary condition ($(\nabla -iA)\Psi|_n=0$) while for a 
superconductor-normal boundary condition 
(approximated by the limiting case of $\Psi|_S=0$) 
$H_p$ has a smaller value  independent of $\kappa$ and 
proportional to $H_c$.
We study the magnetization curves and penetration
fields at different sample sizes and for square and thin film
geometries. 
For small mesoscopic samples we study the peaks and discontinuous
jumps found in the magnetization as a function of magnetic field.
To interpret these jumps we consider 
that vortices located inside the sample induce a 
reinforcement of the surface barrier 
at fields greater than the  first penetration field $H_{p1}$. 
This leads to multiple penetration fields $H_{pi} = H_{p1}, H_{p2},
H_{p3}, \ldots$ for vortex entrance in mesoscopic samples. 
We study the dependence with sample size of the
penetration fields $H_{pi}$. We explain these multiple penetration
fields extending the usual Bean-Livingston analysis by considering
the effect of vortices inside the superconductor and 
the finite size of the sample.

\end{abstract}

]

\section{INTRODUCTION}

In the last years there has been an important experimental and
theoretical interest in the study of vortex physics in a 
mesoscopic scale.\cite{buisson,geim98,bolle,geim00}
The smallness of these systems imply that the sample
geometry and   the interaction between
the vortices and the sample surface become important. 
The interaction between vortices and the surface  manifests
itself fundamentally in the existence of a surface barrier,
first studied by Bean and Livingston,\cite{bean} 
which delays the vortex 
penetration and generates metastable states. If  the
surface effects are ignored, 
the penetration of magnetic field is energetically
favorable at the first critical field $H_{c1}$. However the energy
barrier of the surface 
prevents the vortex entry until a higher field $H_p$ at which the 
barrier vanishes. $H_p$, also known as the superheating
field, is associated with the peak in the 
magnetization curves and is strongly influenced by the presence of 
surface irregularities. 

The surface barrier has attracted a 
renewed interest recently in the study of mesoscopic superconductors.
For example,  Enomoto and Okada \cite{enomoto} by means of 
numerical simulations of the time-dependent Ginzburg-Landau 
equations (TDGL), studied the influence of  
temperature and surface irregularities on the surface barrier.
 Sonin and Traito \cite{sonin} showed that the presence of the 
 surface barrier affects the entry and exit of vortices influencing 
 the surface resistance. They found a surface-induced 
suppression of the ac losses.

One important line of research of mesoscopic superconductors are
the superconducting disks. 
\cite{buisson,geim98,geim00,peeters97,peeters98a,peeters98b,peeters99a,peeters99b,kuznetsov,peeters99c} 
The study of small superconducting disks was started by Buisson {\it et
al.} \cite{buisson} for disks with radius $\sim$ 7 $\mu$m.
Recent advances in the microfabrication technology and
measurement techniques now allow the fabrication and study of
superconducting disks with sizes comparable to the 
coherence length $\xi$, with radius  as small as  0.3
$\mu$m.\cite{geim98} 
Most of the studies were done in Al disks, a 
material with $\kappa \simeq 0.3$, 
however for small samples the effective penetration depth 
$\Lambda =\lambda^2/d$ increases for decreasing disk thickness 
($d$) resulting
in effective $\kappa$ values in the type II region that can be
studied theoretically using the equilibrium Ginzburg-Landau equations. 
In this regime the Al disk can develop Abrikosov multivortex states
\cite{peeters98a} and depending on the radius ($R$) and thickness of 
the disk it is possible to observe first or second order phase 
transitions,\cite{peeters97} by increasing the disk 
sizes the second order reversible phase transition observed for 
small disk radius is replaced by a first order transition. 
There is also an intermediate regime where jumps in the 
magnetization appear associated with the vortex entrance.
Other interesting phenomena have been studied for
mesoscopic Al disks, for example in Ref.\onlinecite{peeters99c} 
hysteresis in the magnetization curves was observed experimentally 
and explained in terms of the presence of a ``Bean-Livingston''
surface barrier and in Ref.~\onlinecite{peeters99b} 
the behavior of the third critical field $H_{c3}$ was investigated 
for different sample sizes and geometries.

The time dependent Ginzburg-Landau (TDGL) equations have been 
proposed \cite{tdgl0} as a  time dependent generalization
of the mean field approach of the Ginzburg-Landau theory.
Gorkov and Eliasberg \cite{gorkov} obtained the
TDGL equations from the microscopic BCS theory in the gapless case. 
In the last years, numerical simulations of the time dependent 
Ginzburg-Landau (TDGL) equations 
have been suscesfully used to study
the magnetic properties and flux dynamics in superconductors.
\cite{frah,liu,kato,machida,aranson,alvarez}
Frah {\it et al.} \cite{frah} and Liu {\it et al.} \cite{liu} simulated
the TDGL equations for $\kappa=0.3-20$, Kato {\it et al.} \cite{kato}
and Machida and Kaburaki\cite{machida} for $\kappa=2$, 
Aranson {\it et al.} \cite{aranson} studied vortex 
dynamics in the $\kappa=\infty$ limit , and Vicente-\'{A}lvarez {\it et
al.} studied the dynamics of $d$-wave superconductors.\cite{alvarez}
Only few studies of the TDGL equations
have been done in superconductors with type I behavior,\cite{frah,liu}
possibly because the theory is better to describe a 
superconductor near a second order phase transition 
at a temperature near $T_c$.

In this paper, we present a numerical simulation of the TDGL 
equations to study the surface barrier in mesoscopic samples
for $\kappa=0.15-2$. We neglect demagnetization effects and therefore we assume
that the sample is infinite in the direction of the external magnetic
field (the ${\bf \hat z}$ direction). We consider square samples that
are mesoscopic in the $xy$ plane (perpendicular to the magnetic field)
with linear sizes of $5-30 \lambda$, with $\lambda$ the penetration
depth. 

The paper is organized as follows. In Sec.II we present the 
TDGL equations with their discretized form in finite differences,
and we discuss their possible boundary conditions.
In the Sec.III of the paper we study the dependence 
of the penetration field
$H_p$ with the Ginzburg-Landau parameter $\kappa$ 
($\kappa =\lambda /\xi$), exploring both the 
type I ($\kappa < 1/ \sqrt{2}$) and the type II region 
($\kappa > 1/ \sqrt{2}$) for large samples. 
We study the effects of the surface barrier from a comparison of
two types of boundary conditions.
(i) The superconductor-insulator (S-I) boundary condition:
consisting in the vanishing of the superconducting current
perpendicular to the boundary (${\bf J_s\cdot {\hat n}}=0$).
In this case we find an increase of 
$H_p$ with decreasing $\kappa$. 
(ii) The superconductor-normal (S-N) boundary condition:
approximated as the vanishing of the 
superconducting order parameter at the boundary
($\Psi|_S=0$). A different behavior is
observed for the S-N boundary condition, the field $H_p$ is 
independent of $\kappa$ and nearly equal to $H_c$.   
In the Sec.IV we study the surface barrier in mesoscopic
superconductors. In particular, in Sec.IVA
we study magnetization curves in
type II superconductors at different sample dimensions in the 
region where the transition 
from a macroscopic to a mesoscopic behavior takes place. 
In Sec.IVB we show that the discontinuities that appear in the 
magnetization curves of mesoscopic samples can be 
explained by considering that the vortices that are inside the 
sample induce a reinforcement of the surface barrier at fields 
greater than the first penetration field. In this way, 
it is possible to define a second, third, fourth, etc.  penetration 
fields which are a consequence of the interaction between vortices 
and the surface currents. We study the sample size dependence of 
the first, second and third penetration fields and  
we show that for sufficiently 
large sample sizes the known macroscopic behavior is recovered, 
i.e. a continuous magnetization curve appears since 
$H_{p3} \rightarrow H_{p2} \rightarrow H_p$.
Finally in Sec.V we give a summary of our results and conclusions.
 
\section{MODEL AND DYNAMICS}

\subsection{TDGL equations}

Our numerical simulations are carried out using the 
time-dependent Ginzburg-Landau equations complemented with the 
appropriate Maxwell equations. In the zero-electric potential 
gauge we have: \cite{tdgl0,gropp} 
\begin{eqnarray} 
\frac{\partial \Psi}{\partial t} = \frac{1}{\eta} [(\nabla -
iA)^2 \Psi +(1-T)(1-|\Psi |^2)\Psi ]   \\
\frac{\partial A}{\partial t} = (1-T)\mbox{Im}[\Psi^* (\nabla -
iA)\Psi] -\kappa^2\nabla \times \nabla \times A
\end{eqnarray}
where $\Psi$ and $A$ are the order parameter and vector potential
respectively and  $T$ is the temperature.\cite{noise} 
Equations (1) and (2) are in their dimensionless form.
Lengths have been scaled in units of $\xi(0)$,
times in units of
$t_0=4\pi \sigma\lambda^2/c^2=\xi(0)^2 /\eta D$, 
$A$ in units of $H_{c2}(0) \xi (0)$ and
temperatures in units of $T_c$. $\eta$ is proportional to the ratio
of characteristic times for $\Psi$ and $A$,
$\eta=t_\Psi/t_0=c^2/(4\pi\sigma\kappa^2D)$, 
with $t_\Psi=\xi^2/D$, where $\sigma$ is the quasiparticle conductivity 
and $D$ is the
electron diffusion constant. For superconductors with magnetic impurities
we have $D=c^2/(48\pi\kappa^2\sigma)$, and therefore $\eta=12$ in this case.

We have used the standard finite difference 
discretization scheme to solve equations (1) and (2). \cite{gropp} 
The order parameter and vector potentials are defined
at the nodes of a rectangular mesh ($\vec{r}=(I,J)$), and the 
link variables $U_{_\mu I,J}=\exp(-\imath \kappa h_{\mu} A_{\mu I,J})
\;\;\; (\mu=x,y)$ are introduced in order to maintain the gauge 
invariance under discretization.

In our simulations we have assumed a sample that  
has a square/rectangular shape in the $x,y$ 
direction with dimensions $L_x\times L_y$ 
and it is infinite in the $z$ direction. 
We apply the magnetic field parallel to the $z$ direction, 
the symmetry of the problem  then implies for
 all mesh points $A_{I,J}=(A_{xI,J},A_{yI,J},0)$ and 
 ${\bf B}_{I,J}=(0,0,B_{zI,J})$, where 
  $B_{_zI,J}=(\nabla \times
  \vec{A})_z=(\partial_xA_{_yI,J}-\partial_yA_{_xI,J})$.
  
In this geometry the discretized form of equations (1) and (2) are:
\begin{eqnarray} \nonumber
\frac{\partial \Psi}{\partial t}&=&
\frac{1}{\eta}\Bigl(\frac{U_{_xIJ}\Psi_{I+1,J}-2\Psi_{I,J}
+U_{_xI-1,J}\Psi_{I-1,J}}{(\Delta x)^2}+ \nonumber \\
&+&\frac{U_{_yI,J}\Psi_{I,J+1}-2\Psi_{I,J}+U_{_xI,J-1}\Psi_{I,J-1}}{(\Delta
y)^2} + \nonumber \\
&+&(1-\frac{T}{T_c})(1-|\Psi_{I,J}|^2)\Psi_{I,J}\Bigr) \\ 
\frac{\partial A_{_xI,J}}{\partial t}&=&(1-\frac{T}{T_c})\frac{\mbox{Im}
[U_{_xI,J}\Psi^*_{I,J} \Psi_{I+1,J}]}{\Delta x} - \nonumber \\
&-&\kappa^2(\frac{B_{_zI,J}-B_{_zI,J-1}}{\Delta y})  \\
\frac{\partial A_{_yI,J}}{\partial t}&=&(1-\frac{T}{T_c})\frac{\mbox{Im}
[U_{_yI,J}\Psi^*_{I,J} \Psi_{I,J+1}]}{\Delta y} - \nonumber \\
&-&\kappa^2(-\frac{B_{_zI,J}-B_{_zI-1,J}}{\Delta x}) 
\end{eqnarray}
where $\Delta x$ and $\Delta y$ are the mesh widths of the spatial 
discretization.

\subsection{Boundary conditions}

The dynamical equations must be completemented with the appropriate boundary
conditions for both the order parameter and the vector potential. 

The boundary conditions for the vector potentials $A_{\mu I,J}$ 
are obtained by making $$B=\nabla \times A=H_a$$
at the sample surface (where $H_a$ is the applied magnetic field). 

The boundary conditions for the order parameter depend sensibly on the
physical nature of the boundary. In general is given by: \cite{gennes2}
\begin{equation} 
(\vec{\Pi} \Psi)|_n =(\nabla -i\vec{A})|_n \Psi =\frac{\Psi}{b} 
\end{equation}
where $b$ is a surface extrapolation length which embodies the
surface suppression (or enhancement if $b<0$) 
of the superconducting order parameter. 

For the boundary between a superconductor and an insulator 
(or the vacuum) we have $b \sim \xi^2(0)/a$, with $a$ the interatomic
distance.\cite{gennes2} 
For low temperature superconductors $b$ is huge ($\sim 1$cm),
and the superconductor-insulator (S-I) boundary condition is usually
approximated with the limit $b \rightarrow \infty$:
\begin{equation} 
(\nabla-i\vec{A})|_n \Psi =0 
\end{equation}
This boundary condition implies that the perpendicular
component of the superconducting current is equal to zero at the 
surface ($\vec{J}_s|_n=0$). This is the most frequently 
used boundary condition  because it also minimizes the 
free energy at the sample surface. More precisely, this boundary
condition is valid for superconductors with interfaces 
for which $b\gg\xi(T)$.

For the boundary between a superconductor and a normal metal
we have $b\sim \frac{N}{N_N}\frac{1}{T_j}\frac{\xi(0)^2}{\xi_N}$,
with $N$ the local density of states at the Fermi surface, $N_N$ the
bulk density of states, 
$T_j$ the transmission coefficient at the boundary and $\xi_N$ is the
coherence length in the normal state. \cite{gennes2} Typically
$(N_N/N)T_j\sim 1$ and $b$ is small when compared to $\xi(T)$.
Therefore, the superconducting-normal (S-N) boundary condition can
be approximated by $b\approx0$, giving:
\begin{equation}
\Psi|_S =0\;.
\end{equation} 
The case of $b \rightarrow 0$ is also found for
a ferromagnet-superconductor surface.\cite{indeku} Moreover,
the condition $\Psi|_S=0$ is similar to having a high density of defects 
at the interface and therefore a highly defective surface is also
represented by (8).
It is also interesting to note that for high $T_c$ superconductors
even the superconductor-insulator boundary is better approximated by (8)
since $b\sim \xi(0)^2/a \ll \xi(T)$ in a large range of temperatures,
due to the smallness of $\xi(0)$ in this case.   
The boundary condition of (8) completely suppresses the currents
at the boundary 
($\vec{J}_s^\perp=\vec{J}_s^\parallel=0$) and
maximizes the surface Helmholtz free
energy, which becomes equal to the free energy of 
a normal metal. 

In the study of  the surface 
barrier in Sec.III we will compare these two conceptually 
different boundary conditions of the order parameter.
They represent the two  limiting cases of Eq.~(6) and we will
call them, in short, the ``S-I'' boundary condition (Eq.~(7)) and
the ``S-N'' boundary condition (Eq.~(8)).
A previous discussion of these two types of boundary conditions
was done by Buisson {\it et al.} \cite{buisson}, 
where the equilibrium solutions
and eigenvalues of the linearized Ginzburg-Landau equations were
compared to study the behavior near $H_{c2}$. 
However, for the vortex nucleation process at low 
magnetic fields the nonlinear terms
of the Ginzburg-Landau equations should be considered.

\section{SURFACE BARRIER IN MACROSCOPIC SAMPLES}

When the magnetic field $H$ is increased starting from $H=0$ in a finite 
sample, the Meissner state is destroyed at a magnetic
field $H_p$ which is typically higher than $H_{c1}$. 
This is due to the presence of a surface barrier for vortex entrance
in finite samples. The surface barrier in
macroscopic samples is sometimes known as the ``Bean-Livingston barrier''
since it was first studied by  Bean and Livingston (BL) \cite{bean}.
In the BL work the surface barrier 
was obtained in the London approximation and 
for the ideal case of a semi-infinite
superconductor, therefore vortex core nucleation effects and
geometrical effects were both neglected.
With this simplification, the BL  work was
able to identify one of the most relevant causes of the
surface barrier: the screening currents
near the surface circulate in the opposite direction to the
superconducting currents around a vortex. This is typically
viewed as a competition between the  repulsion
between the vortex and the surface shielding current and the
attractive force between a vortex inside the sample and its image.
This argument, which is based on the London model, yields the
Bean-Livingston result for the penetration field, 
$H_p^{{\rm London}} \approx \frac{\sqrt{2}}{2} H_c$, 
which is independent of $\kappa$.\cite{bean,gennes2}
However, in addition to overcoming the barrier induced by the 
shielding Meissner currents, a vortex penetration event has also to deplet the
superconducting order parameter $|\Psi|$ at the surface, resulting 
in a higher penetration field $H_p(\kappa) > H_p^{{\rm
London}}$.\cite{ginzburg,matricon,kramer1,fink,kramer2,dolgert}
This later effect can be sensitive to the boundary conditions of
the Ginzburg-Landau equations.  In this section we will discuss
the effect of boundary conditions on the surface barrier in macroscopic
samples as a function of $\kappa$.

\begin{figure}[htb] 
\centerline{\epsfxsize=8.5cm \epsfbox{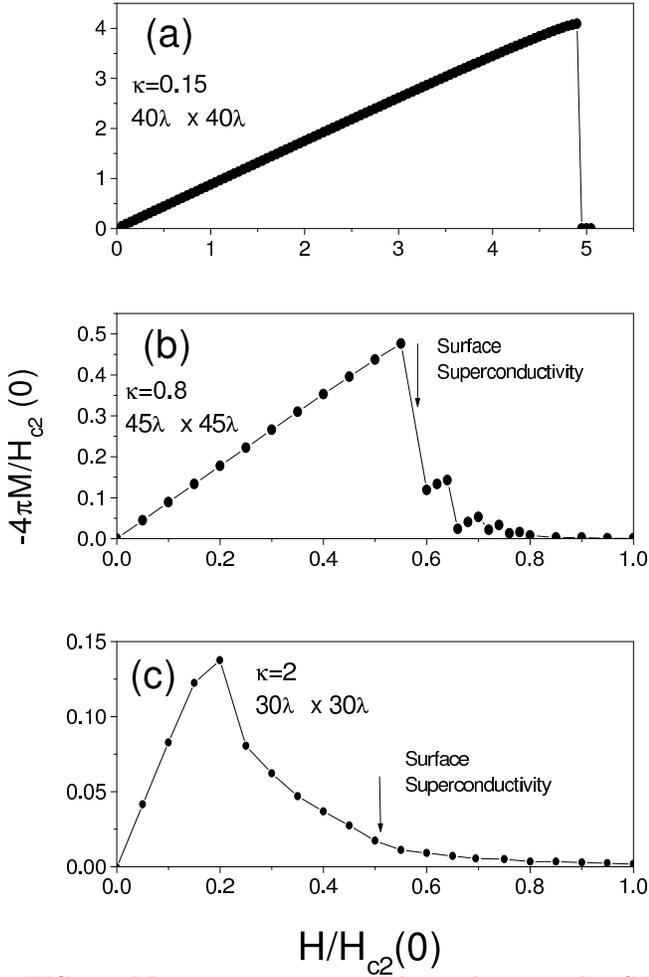}}
\caption{Magnetization curves obtained 
using the S-I boundary condition: (a) $\kappa =0.15$, (b) $\kappa =0.8$
and (c) $\kappa =2$. (Inserted in the figure are the size of the 
superconducting region used in the
simulation, $\lambda$ is the penetration length).} \label{fig1} 
\end{figure}\noindent

We start our study of the surface barrier by first
analyzing the magnetization curves in large samples.
The magnetization curves for the S-I boundary condition are
summarized in  Fig.1 and for the S-N boundary condition are 
summarized in Fig.2. 
In both cases the curves were obtained initializing the
variables to a perfect Meissner state [$\Psi(I,J,t=0)=1$ and
$A(I,J,t=0)=0$] and increasing the magnetic field at subsequent steps,
usually with  $\Delta H=0.05 H_{c2}$. We take as the initial condition 
at a magnetic field $H$ the final state of $\Psi(I,J)$ and $A(I,J)$ of the
previous magnetic field value $H-\Delta H$. In this way we mimic the 
experimental procedure of increasing the magnetic field in a
magnetization measurement. For each magnetic field we  
calculate the magnetization $M$ taking time averages of the
time dependent magnetization $M(t)$:
\begin{eqnarray}
4\pi M(t)&=&\frac{\Delta x\Delta_y}{L_xL_y}\sum_{I,J} 
B_z(I,J,t) - H_a \nonumber\\
M&=& \frac{\Delta t}{t_f-t_o} \sum_{t=t_0}^{t_f} M(t)\;,
\end{eqnarray}
\begin{figure}[htb] 
\centerline{\epsfxsize=8.5cm \epsfbox{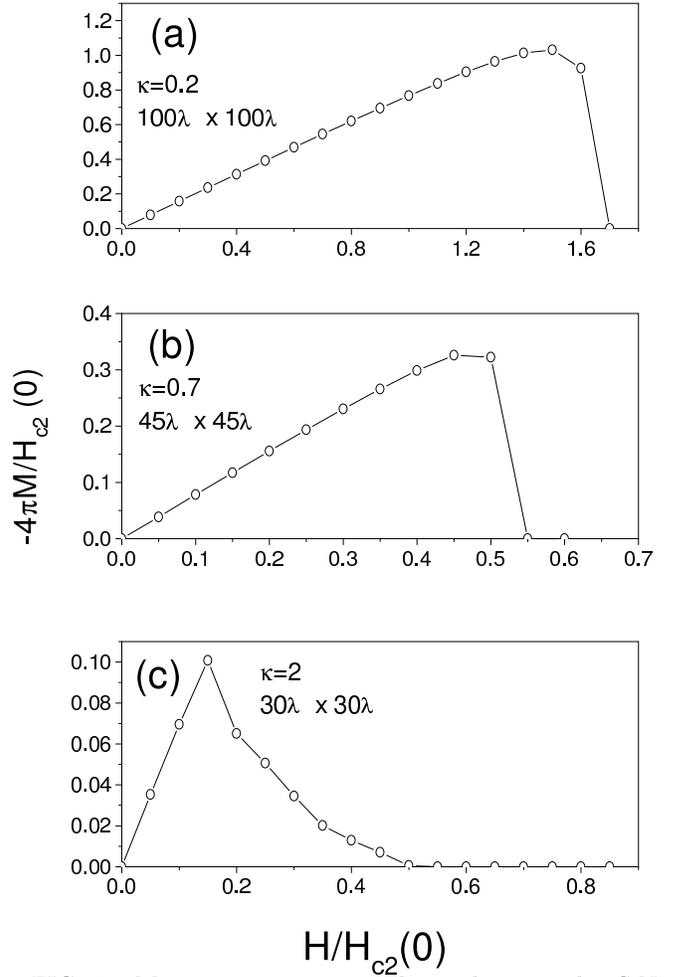}}
\caption{Magnetization curves obtained 
using the S-N boundary condition: (a) $\kappa =0.2$, (b) $\kappa =0.7$
and (c) $\kappa =2$.} \label{fig2} 
\end{figure}\noindent
where we start the average at a time $t_0$ after a steady state
was reached. 
In the simulations we have taken $T=0.5$, 
$\eta =12$ and we used a mesh of $120 \times 120$ 
points. 
In order to make efficient calculations we have chosen the 
time step ($\Delta t$) and the spatial 
discretization ($\Delta x$ and $\Delta y$)
depending on the value of $\kappa$. For example, for 
$\kappa=0.15$ we used 
$\Delta x=\Delta y=0.05$ and $\Delta t =0.0025$; for 
$\kappa =0.8$, $\Delta x=\Delta y=0.3$ and $\Delta t =0.0025$ and
for $\kappa =2$, $\Delta x=\Delta y=0.5$ and $\Delta t =0.015$.
Since the TDGL equations considered here represent
a mean field dynamics,\cite{tdgl0,gorkov,frah,liu,machida,aranson,gropp} 
the effect of thermal noise fluctuations
beyond mean field are neglected,\cite{noise}  which is 
correct for low $T_c$ superconductors. 

Figure 1(a) shows the case of  a type I superconductor with
$\kappa =0.15$. We can see that the TDGL equations 
reproduce the basic phenomenology of type I 
superconductivity characterized by a first-order magnetic transition. 
In  this case the superconductivity disappears 
abruptly and there is  no surface superconductivity. 
The field profile is described by a Meissner state and $H=H_a
\exp(-x/\lambda)$. In this simulation the
intermediate state structures typical of type I superconductors are not
found at equilibrium\cite{nota} since the long range 
interactions between currents and demagnetization effects are
neglected.\cite{goldstein,bokil}
Results for a type II superconductor with $\kappa=2$ 
are shown in Figure 1(c). 
In this case the superconductor is in the
Meissner state until a penetration field $H_p$ is reached. 
At $H_p$ some vortices enter 
the sample and a peak in the magnetization curve is observed. 
Above $H_p$ the magnetization increases due to the penetration of 
vortices in the sample, 
until the establishment of surface superconductivity 
for fields in the range $H_{c3}(T)>H>H_{c2}(T)$. The criterion for 
surface superconductivity  that we use is the existence
of superconductivity in the surface 
(i.e. $|\Psi|\not=0$ in a contour around the surface
of width  equal to the discretization length, $\Delta x = 0.5\xi$ in this case)
and the exact vanishing  of superconductivity in the bulk
(i.e. $|\Psi|=0$ inside the region enclosed  by the surface contour). 

The S-N boundary condition leads to a different magnetic behavior
as can be observed in Figure 2 where we have used similar
parameters as those reported in Figure 1. 
We observe  the following differences: 
(i) The magnetization is smaller for the same 
external magnetic field (there are less vortices). 
Since in this case the superconducting order parameter vanishes
at the surface, the Meissner shielding currents are nucleated at a
distance of a few $\xi$ inside the sample, 
instead of being right at the boundary. 
Therefore less vortices can stay inside the sample
for a given magnetic field when compared with the S-I boundary.
For example, for $\kappa=2$, we find that
the shielding distance $\delta$ between the vortices and the sample
surface is $\delta_{SN}\approx 5\lambda$ for the S-N interface
while for the S-I interface we have $\delta_{SI}\approx 3\lambda$.
In the general case, when the boundary condition is described by  
Eq.(6) with a finite $b$, one can expect that 
the shielding distance will have an intermediate value
$\delta_{SN}>\delta(b)>\delta_{SI}$.
(ii) There is no surface superconductivity above $H_{c2}$. This
is an obvious and direct consequence of the S-N 
boundary condition that enforces $\Psi|_S =0$. 
(iii) The first penetration field $H_p$ is
smaller and therefore the surface barrier is lower,
we will discuss this in detail in the following paragraphs.

In order to understand the difference in the surface barrier,
let us study the dynamics of the first 
vortex penetration just at $H=H_p$.  Figure 3 
shows the dynamic evolution of the order parameter near 
a small region close to
an S-I interface (Figure 3 left) and close to an S-N interface (Figure 3 right). 
If the interface is of the S-I type 
the order parameter at the boundary is different from zero 
in the Meissner state. When the condition for vortex entrance 
is fulfilled the order parameter at a boundary point 
has to decrease until reaching zero. 
Therefore there is an intermediate time
interval when $\Psi=0$ in a point at the boundary. 
Just in this moment a vortex can enter the sample  and
afterwards the order parameter at the boundary increases
again and returns to a non-zero value. It is interesting to note that this 
process is always necessary for vortex penetration
in S-I interfaces, there is always the need of an intermediate 
$\Psi=0$ state at the boundary. 
On the contrary, a smooth entrance of vortices is observed for 
an S-N interface (Figure 3 right). 
\begin{figure}[htb] 
\centerline{\epsfxsize=8.5cm \epsfbox{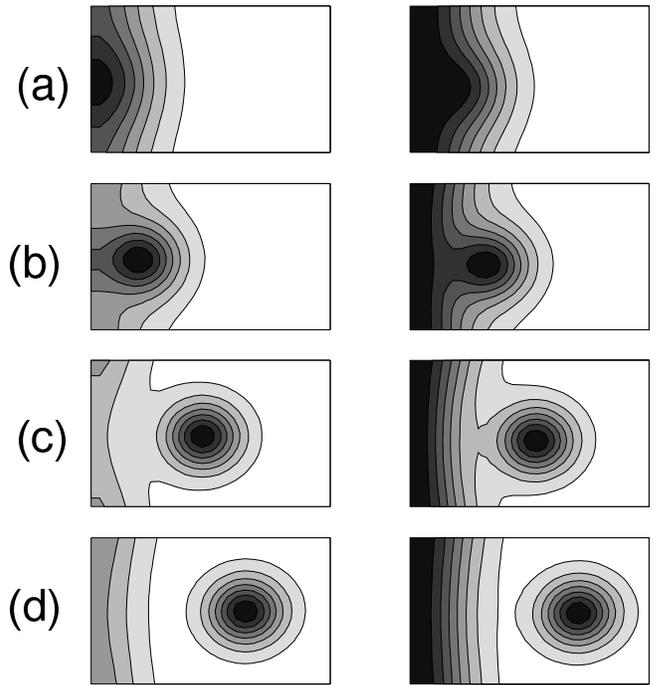}}
\caption{Time evolution of the spatial pattern of the order parameter in a small 
region around the boundary were a vortex entrance is taking place. For 
the S-I condition (left) the magnetic field is 
$H= 0.199 H_{c2}(0)$, and for the S-N condition (right) $H= 0.169 H_{c2}(0)$. In both
cases $H$ is just above the first penetration field and $\kappa=2$. Gray scale from
black ($|\Psi|=0$) to white ($|\Psi|=1$).} 
\label{fig3} 
\end{figure}\noindent
For this boundary condition the order 
parameter at the interface is already zero. A small deformation
of the region of $\Psi=0$ allows for the penetration of a vortex. 
Since there is no need of depressing the order parameter at the
boundary, the surface barrier is much smaller in the S-N case.
A similar dynamical behavior for vortex entrance would appear 
in an S-I interface with defects at the surface. 
At the defects the order parameter is depressed and therefore
$\Psi=0$ is already established at some boundary points,
which are preferred points for vortex entrance.
 
From the magnetization curves  
we can obtain the first field for flux penetration $H_p$ as a 
function of $\kappa$ for the different boundary conditions, this is
shown in Figure 4.
The $\kappa$ dependence of the superheating field $H_p$ 
has been previosly calculated
for the case of a semi-infinite medium  
with the Ginzburg-Landau
equations.\cite{ginzburg,matricon,kramer1,fink,kramer2,dolgert}
Matricon and Saint James\cite{matricon} obtained $H_p(\kappa)$ solving
the semi-infinite one-dimensional GL equations,
in Ref.~\onlinecite{kramer1,fink,kramer2,dolgert}  
the stability of the superheated state under small fluctuations of 
the order parameter and the vector potential was discussed,
and very recently Vodolazov \cite{vodolazov}  
analyzed the effects of surface defects on $H_p$. 
In the case of a one dimensional 
semi-infinite medium,
the Matricon-Saint James\cite{matricon} solution can be obtained solving 
the equations:
\begin{eqnarray}
\frac{d^2\Psi}{\kappa^2 dx^2} &=& A^2 \Psi + \Psi - \Psi^3, \nonumber \\
\frac{d^2A}{dx^2}&=&\Psi^2A
\end{eqnarray}
with the boundary conditions:
\begin{eqnarray}
H=H_a  \;\; \mbox{and} \;\; \frac{d\Psi}{dx}=0 \;\; &\mbox{at}& \; x=0 \nonumber \\
A=H=0  \;\; \mbox{and} \;\; \Psi =1 \;\; &\mbox{at}& \; x=\infty 
\end{eqnarray}
Solving numerically equations (9) with the boundary condition (10)
it is possible to find a relationship among $H_a$ and $\Psi(x=0)=\Psi_o$,
where $H_{p}$ is the maximum value of 
$H_a$ that allows a physically meaningful $\Psi_o$, i.e.
$0<\Psi_o <1$. The results obtained in this way are represented 
by the continuous-line of Figure 4. Our simulational results, on the 
other hand, are a numerical solution of the exact problem in a 
two dimensional square sample (which has geometrical effects) 
and are represented  by closed circles. 
The $H_p$ values reported here are the
peaks of the simulated dc-magnetization curves 
and the error bars correspond to the discrete field step used 
in the the magnetization curves. We see in Fig.~4
that the value of $H_p$ obtained for the S-I interface
is always well above $H_{c1}$
in the type II region ($H_{c1}(T)=[(\ln\kappa)/\sqrt{2}\kappa]H_c(T)$)
and above $H_c$ in the type I region, supporting the existence 
of a ``surface barrier'' even in the type I case.  
Our numerical simulations show that for the S-I interface
$H_p$  increases with decreasing $\kappa$  
and $H_p(\kappa) > H_c$, with a behavior
in good agreement with the result for the semi-infinite medium 
obtained by Matricon-Saint James. 
(However, for smaller mesoscopic samples the value of $H_p$ can be
significantly larger than the Matricon-Saint James result, enhancing
the geometrical effect of the square sample, see Sec.~IV). 
For $\kappa \rightarrow 0$ our results agree with 
the known  behavior of the 
superheating field in type I supeconductors, $H_p \sim 1/\sqrt{\kappa}$.
\cite{ginzburg,matricon,kramer1}
For $\kappa \rightarrow \infty$ 
we obtain that $H_p \rightarrow H_c \;$
in agreement with the result of Ref.~\onlinecite{ginzburg,matricon,kramer1}. 
Figure 4 also shows that in the type II region the $H_{c2}$ values 
obtained from the S-I simulations (closed squares) are close to the 
expected values ($H_{c2}(T)=\sqrt{2}\kappa H_c(T)$, dotted line). 
Some differences in $H_{c2}$ appear at 
small $\kappa$ near the type I region, since 
in this  region the field
$H_p$ is close to $H_{c2}$ and a delay of the superconducting-normal 
transition could be expected. 

The behavior of $H_p$ vs $\kappa$ is very different when the S-N boundary 
condition is used. 
In this case, $H_p$ is independent of $\kappa$ 
and nearly equal to 
$H_p^{{\rm London}}\propto H_c$ 
(see the open circles of Figure 4).
This result shows that in the case of the S-N interface,
when the condition $\Psi|_S=0$ is enforced, the surface barrier
is only due to the surface shielding currents and  well described
by the London approximation value of Bean and Livingston: 
$H_p^{{\rm London}}$.\cite{bean,gennes2} 
On the other hand, in the case of the S-I interface the penetration field
$H_p$ is higher due to the  extra contribution needed 
for the vanishing of the order parameter at the surface
in a vortex penetration event.
In a type I superconductor, the ``surface barrier'' can
be interpreted as the barrier for penetration of the normal state from
\begin{figure}[htb] 
\centerline{\epsfxsize=8.5cm \epsfbox{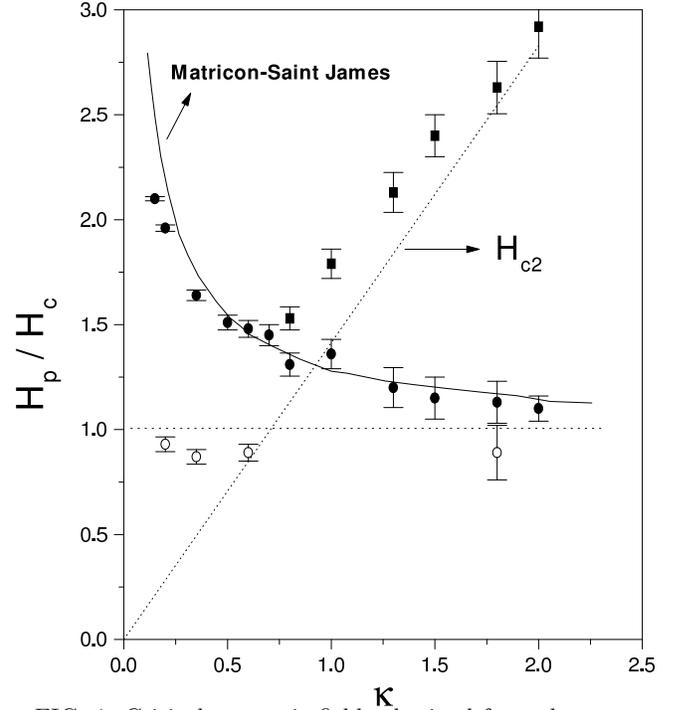}}
\caption{Critical magnetic fields obtained from the
magnetization curves at different $\kappa$ values. For the S-I
boundary condition are plotted the superheating field ($H_p$) 
(closed circles) and the second critical field ($H_{c2}$) (closed
squares). For the S-N boundary condition is plotted $H_p$ (open 
circles). The dashed and dotted curves
are the expected values of $H_c$
and $H_{c2}$ respectively. The continuous line is the result of
$H_p \; vs \; \kappa$ obtained by Matricon and Saint-James for a
semi-infinite sample.} 
\label{fig4} 
\end{figure}\noindent
the boundary. In the S-N interface of a type I superconductor
there is no barrier, {\it i.e.}
$H_p=H_c$, which is an obvious result since the boundary condition
already enforces the normal phase ($\Psi=0$) at the surface. 
On the contrary, in the
S-I interface of a type I superconductor
 the barrier for nucleation of the normal phase at the
boundary can be very high, $H_p \gg H_c$, as can be seen in Fig.4.

In the general case, when the boundary condition of a superconducting sample
is described by  Eq.(6) with a finite $b$, 
we expect that $H_p(b,\kappa)$ will be in between the two limit
cases studied here, $H_p^{SN}(b=0,\kappa)< H_p(b,\kappa) <
H_p^{SI}(b=\infty,\kappa)$.
Ideally, most of the superconductors should be closer to the 
behavior of $H_p^{SI}(b=\infty,\kappa)$ but the effect of
a finite $b$ and the presence of defects at the surface will
give a smaller value of $H_p$ with a lower bound given by 
$H_p^{SN}(b=0,\kappa)$.

\begin{figure}[htb]
\centerline{\epsfxsize=8.5cm \epsfbox{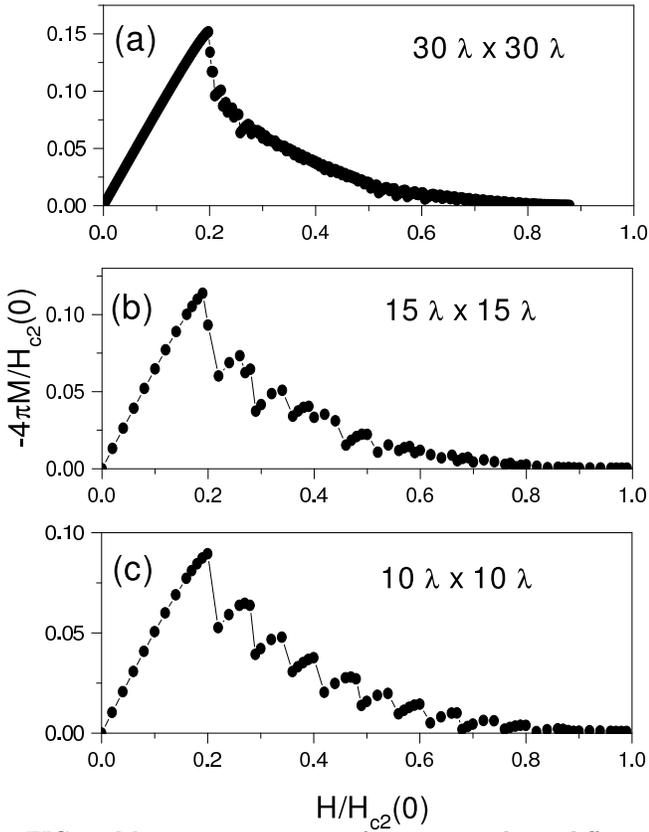}}
 \caption{Magnetization curves of square
samples at different sizes and using the S-I boundary condition  for
$\kappa=2$
(a) $30 \lambda \times 30 \lambda$, (b) $15 \lambda \times 15 \lambda$ and
(c) $10\lambda \times 10 \lambda$.} 
\label{fig5} 
\end{figure}\noindent

\section{SURFACE BARRIER IN MESOSCOPIC SAMPLES}

\subsection{Finite-size effects in mesoscopic type  II
superconductors}

The magnetic behavior of mesoscopic superconductors is different
from the behavior of bulk samples. In the mesoscopic scale, several
maxima appear in the magnetization curves which are related to 
the vortex entrance. This result is quite general and appears 
either in thin films at parallel fields \cite{bolech} or 
mesoscopic superconducting disks. \cite{peeters97,peeters99b} 
In this section we study 
the magnetic behavior of superconducting samples of different 
sizes, in particular we cover the sample size region where a 
transition from a mesoscopic to a bulk behavior takes place. 

Figure 5 shows the dc-magnetic behavior of superconducting square
samples of different sizes with S-I boundary condition. 
The behavior of Fig.5(a) is
typical of the macroscopic samples that we have studied in the
previous section, however if we decrease the sample size to
the mesoscopic region the continuous behavior disappears and some
magnetization maxima followed by discontinuous jumps appear 
(see Figs.5(b) and (c)). 
\begin{figure}[htb]
\centerline{\epsfxsize=8.5cm \epsfbox{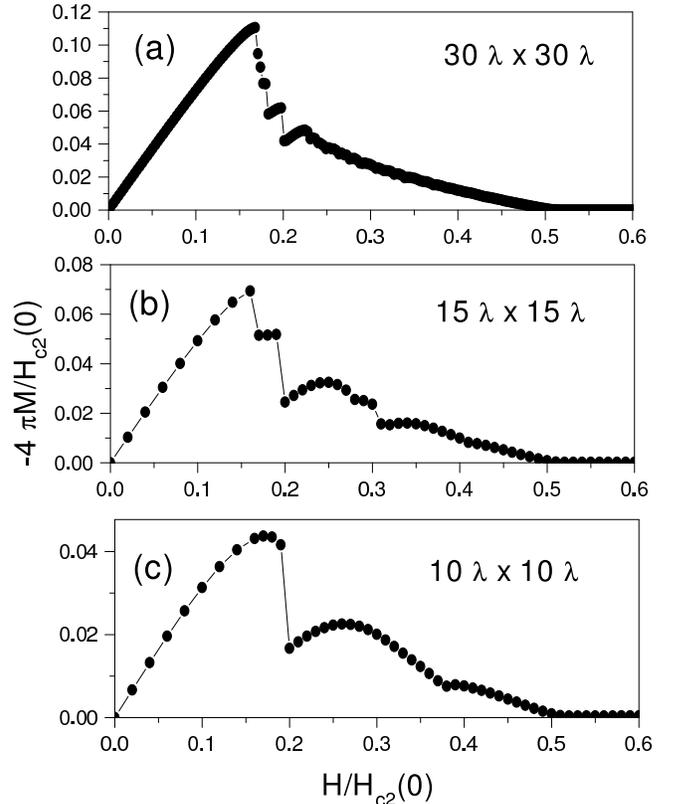}}
\caption{Magnetization curves for the S-N
boundary condition and $\kappa=2$. We show the magnetic curves at different 
sample sizes from the macroscopic to the mesoscopic-like behavior: 
(a) $30 \lambda \times 30 \lambda$, (b) $15 \lambda \times 15 \lambda$ 
and (c) $10\lambda \times 10 \lambda$.} 
\label{fig6} 
\end{figure}\noindent
We find that the
discontinuities in the magnetization correspond to the
penetration of new vortices into the sample, this will be discussed
in detail in Sec.IVB. 

If we change the boundary condition and we use the `S-N' 
condition we find, as it is shown in Figure 6, 
a different mesoscopic behavior: there is an 
appreciable decrease in the number of magnetization maxima and 
the transition between the states of the system with a different 
number of vortices seems more continuous. 
In other words, for the same sample size, 
fewer vortex penetration events are needed to arrive to the normal state.
The decrease in the number of
entrance events for the S-N boundary in mesoscopic samples is 
related to the fact that this boundary condition allows less
vortices inside the sample, as discussed in the previous section.
The fact that the S-N interface needs a larger shielding distance,
$\delta_{SN} > \delta_{SI}$, has clearly stronger consequences 
in the magnetic behavior of a mesoscopic sample.
At the same magnetic field
the mean magnetization values of the Meissner state are lower in 
the S-N case than in the S-I case as can be observed comparing 
Figures 5(c) and 6(c). 
In general a mesoscopic sample with a boundary corresponding
to $0<b<\infty$ in Eq.(6), will have a magnetization behavior
in between the two limiting cases of Fig.~5 and Fig.~6. 

\begin{figure}[htb] 
\centerline{\epsfxsize=8.5cm \epsfbox{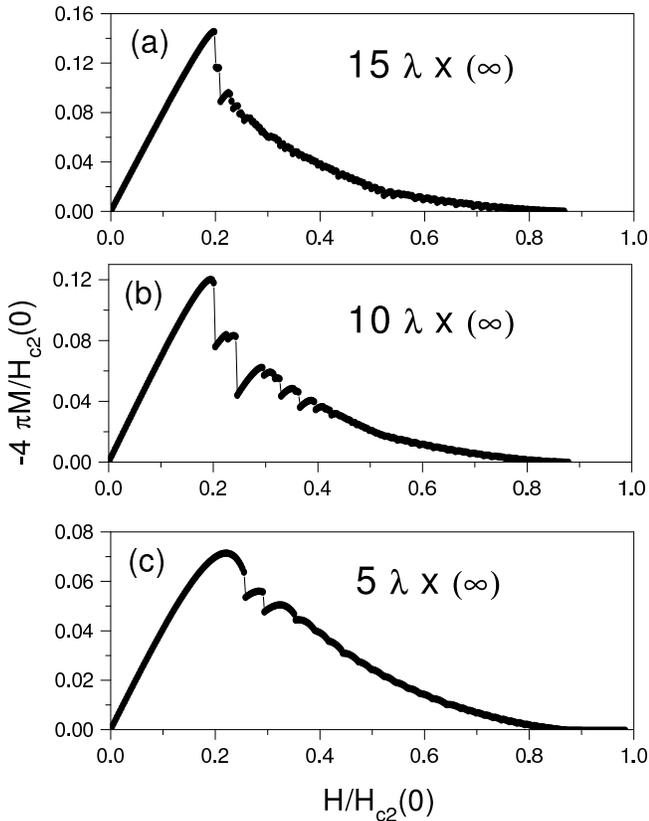}}
\caption{Magnetization curves of thin films
with the external magnetic field applied parallel to the film. The
curves are generated for decreasing thicknesses of the film and using
the S-I boundary condition for $\kappa=2$. 
 (a) $15\lambda \times \infty$, (b) $10 \lambda \times \infty$ 
 and (c) $5\lambda \times \infty$.} 
\label{fig7} 
\end{figure}\noindent

The changes observed in the magnetization curves when
going from the macroscopic to the mesoscopic behavior are
quite general and do not depend on the sample geometry. We also 
study the magnetic behavior of a thin film with the field parallel to
its faces. In this case there is only one relevant dimension, the 
thickness of the film ($d$). We work here in the case 
$d> \lambda$. Figure 7 shows the magnetic behavior of 
thin films with different thickness obtained using the S-I 
boundary condition. We 
observe that in this case the discontinuities in the 
magnetization curves appear at smaller film thickness and that they 
are less important
than in squares samples, possibly because vortices in a mesoscopic 
square sample are more confined than in thin films of the same linear
size.
We have also done numerical calculations of the 
magnetic behavior in thin films 
using the S-N boundary condition (see Figure 8). In
particular Figure 8(c) shows that for very small film thickness
there is a continuous change in magnetization 
without the jagged structure observed in Fig.8(a-b). 
In this case vortex lines do not 
penetrate the sample and the superconducting state disappears 
gradually. This behavior is also present when we use 
the S-I boundary condition but it is necessary to explore lower 
sample sizes than the one shown in the corresponding figures.

\begin{figure}[htb] 
\centerline{\epsfxsize=8.5cm \epsfbox{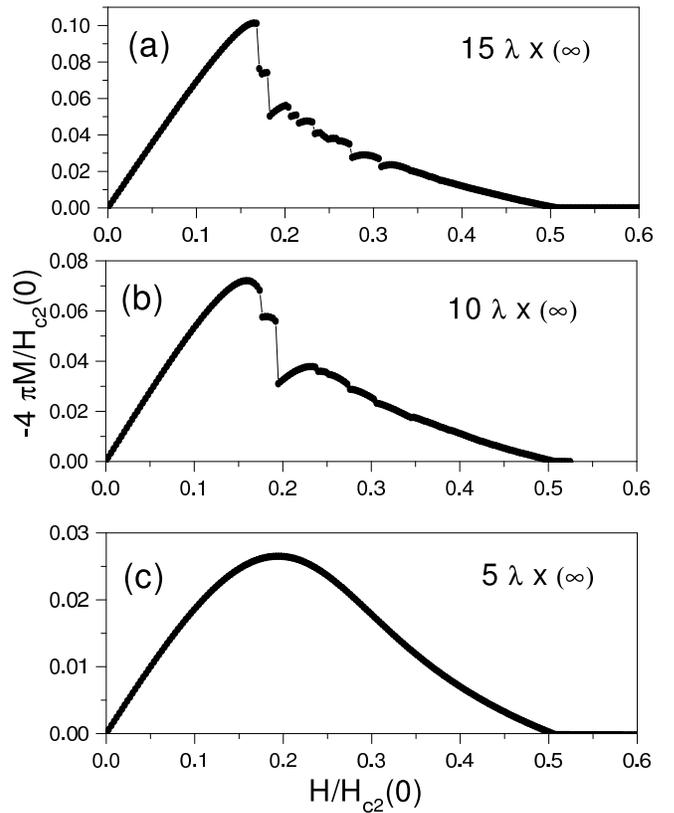}}
\caption{Thin film magnetization curves with
the external field applied in the same geometry of Figure 7 but using
the S-N condition:  (a) $15\lambda \times \infty$, 
(b) $10 \lambda \times \infty$ and (c) $5\lambda \times \infty$. 
As it is observed for samples sizes as low as $5\lambda \times
\infty$, there is a continuous transition to the 
normal state.} 
\label{fig8} 
\end{figure}\noindent

\subsection{Multiple penetration fields in mesoscopic samples}

The behavior of the magnetization in mesoscopic type II samples 
is characterized by maxima followed by discontinuous jumps
as a function of the external magnetic field. 
In Figure 9 we show in detail the case for a sample of size $10\lambda
\times 10\lambda$ and $\kappa=2$.
At the same time we plot the total number of vortices inside the 
sample, $N_v=\frac{1}{\Phi_o}\oint (A+\frac{J_s}{|\Psi|^2})dl$. 
We see that each discontinuous jump in $M(H)$ corresponds
to an increase of $\Delta N_v = 4$ in the number of vortices. 
These jumps occur at succesive magnetic fields 
$H_{pi} = H_{p1},  H_{p2}, H_{p3}, \ldots$, which are indicated
in the figure. In the regions of $H_{pi} < H < H_{p(i+1)}$ the
number of vortices $N_v$ is constant. Therefore the only penetration
events occur at $H_{pi}$, when one vortex enters in each of the
four sides of the square sample. In the region of constant vorticity,
$H_{pi} < H < H_{p(i+1)}$, one may think that no vortices enter
the sample because they feel a surface barrier which is
enhanced by the repulsion force exerted by the vortices 
already inside the sample.

\begin{figure}[htb] 
\centerline{\epsfxsize=8.5cm \epsfbox{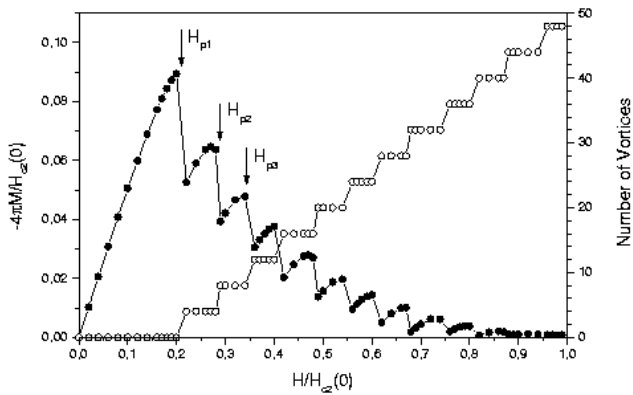}}
\caption{Magnetization curves for a mesoscopic
 $10\lambda\times10\lambda$ square sample
using the S-I boundary condition for  $\kappa =2$. 
In the right scale the number of vortices $N_v$ is shown. 
} \label{fig9} 
\end{figure}\noindent

To analyze this effect, 
let us extend the simple London approximation model of Bean and Livingston
for the surface barrier   
to the case when there are extra vortices inside the sample
and near the boundary. In the London approach, 
the Gibbs free  energy ($G$) of a vortex line located a distance $x$ from 
the sample surface can be calculated as:\cite{bean,gennes2}
\begin{equation}
G=
\frac{\Phi_o}{4\pi}[H_a \exp(-\frac{x}{\lambda})-\frac{1}{2} 
\frac{\Phi_o}{2 \pi \lambda^2} K_o(\frac{2x}{\lambda})+H_{c1}-H_a]
\end{equation}
where $H_a$ is the external magnetic field and $K_o$ is the
modified Bessel function of the second kind. In normalized units
the above expression reads:
\begin{equation}
\frac{4 \pi G}{ H_{c2} \Phi_o}=
[H_a \exp(-\frac{x}{\lambda})-\frac{1}{2 k^2} 
 K_o(\frac{2x}{\lambda})+ \frac{\ln \kappa}{2 \kappa^2}-H_a]
\end{equation}
where we have used the relation $H_{c1}=(\ln \kappa)/2\kappa^2 H_{c2}$.
The first term of Equations (12) and (13) is related with the repulsive
interaction between the vortex line and the external field, the 
second term describes the attraction between the vortex and the 
surface currents, this term is usually interpreted as the 
interaction with an image vortex \cite{bean,gennes2} and
the third term is the vortex self energy. 

The above expressions are valid when there are no vortices 
inside the sample. If there are vortices located inside the
sample, some additional terms due to the interaction between the
vortices are needed. In the following we will assume that there is 
only one vortex inside the sample located at position $l$ and that
we are analysing the Gibbs free energy in the same line perpendicular 
to the sample surface. If $l$ is small, the new vortex that is 
trying to enter
the sample now feels two additional terms. 
The first term is the 
repulsive force between the vortices and the 
second is the attractive interaction with the image of the vortex 
located inside the sample. The last term is needed in order to
take into account the perturbation of the
\begin{figure}[htb] 
\centerline{\epsfxsize=8.5cm \epsfbox{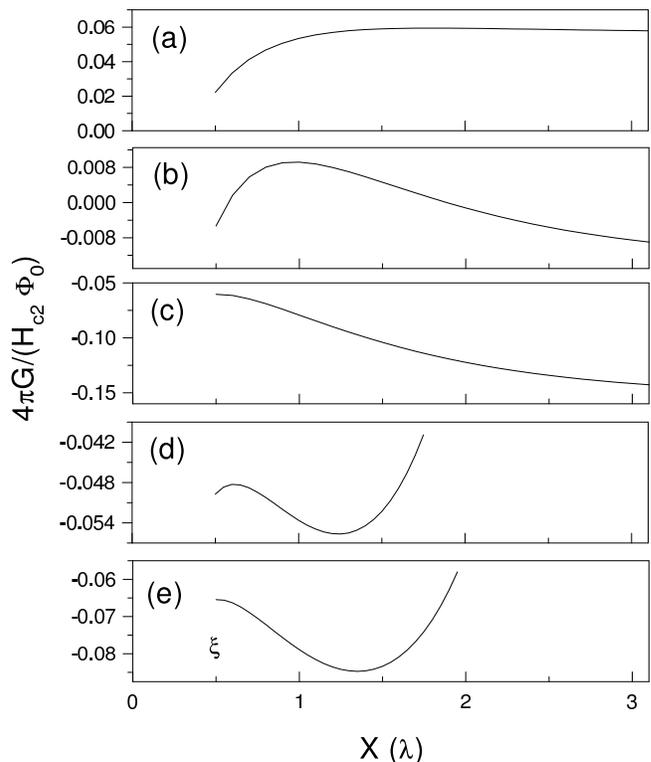}}
\caption{Gibbs free energy of a vortex line as
a function of the distance $x( \lambda)$ from the sample surface. We
have used Equation (12) for (a) $H=0.03 H_{c2}$, (b) $H=0.1 H_{c2}$,
(c) $H=0.24 H_{c2}$ and Equation (13) for (d) $H=0.24 H_{c2}$ and
(e) $H=0.28 H_{c2}$. } 
\label{fig10} 
\end{figure}\noindent 
 vortex already inside
the sample because of its proximity  
to the surface. Both
contributions are more important in mesoscopic superconductors as
we will show below. The free energy that gives the correct force 
expression is:
\begin{eqnarray}
\frac{4 \pi G}{H_{c2} \Phi_o}=
H_a \exp(-\frac{x}{\lambda})-\frac{1}{2 \kappa^2} 
 K_o(\frac{2x}{\lambda})+ \frac{\ln \kappa}{4 \kappa^2}-H_a + \nonumber \\
 + \frac{1}{\kappa^2} K_o(\frac{l-x}{\lambda}) - 
 \frac{1}{\kappa^2} K_o(\frac{l+x}{\lambda})
\end{eqnarray}
Equations (13) and (14) are approximate expressions and have 
the limitation that they are not valid near the sample surface, then 
we will only use it for $x> \xi$. It is useful to note that 
in a more exact treatment we should obtain that $G \rightarrow 0$ 
when $x \rightarrow 0$ because the Gibbs free energy of a vortex 
located outside the sample must be zero.

Figures 10(a), 10(b) and 10(c) were generated using Equation (13) 
for $\kappa=2$. The free energy of the vortex depends on both the 
applied 
magnetic field and the distance to the surface. For $H<H_{c1}$ the 
free energy associated with a vortex is positive at all $x$ values,
as it is shown in Figure 10(a), vortices are then thermodynamically 
unstable inside the superconductor and there is an energy cost 
associated with the vortex entrance. The thermodynamic condition 
for vortex penetration is not fulfilled until $H>H_{c1}$, when 
the free energy of a vortex located well inside the superconductor
becomes negative, see Figure 10(b). However, Figure 10(b) also
shows that because of the attractive interaction between the 
vortex and the surface currents a barrier to vortex entrance 
appears nears the surface. This is the ``Bean-Livingston'' 
surface barrier originated by the screening currents. 
Taking into account 
Equation (13) it is possible to estimate $H_p$ as the magnetic 
field for which the expression 
$( \partial G/ \partial x)_{x= \xi}$ becomes negative, i.e. when 
the maximum of $G(x)$ moves inside the region $x< \xi$. This
condition is fulfilled at $H_p= \frac{\sqrt{2}}{2}H_c$ 
($H_p\approx 0.24 H_{c2}$  in Figure 10(c)). 
At $H=H_p$ some vortices penetrate the 
sample and the free energy associated with the entrance of a new 
vortex now must be calculated using an analogous to Equation (14),
the exact expression depends on the number of vortices and their
location inside the sample. 
Considering only one vortex located at $x=l=3 \lambda$ and using 
Equation (14) we have calculated the free energy just after the first 
vortex entrance ($H>0.24 H_{c2}$), the results are shown in 
Figure 10(d). Observe that the free energy changes considerably
from Figure 10(c) to figure 10(d). Now there are three relevant 
regions: i) near $x=3 \lambda$, there is a region where the Gibbs 
free energy increases with increasing $x$, the strong repulsive 
interaction with the vortex inside the sample dominates; ii) 
there is an intermediate region where G decreases at increasing
$x$, this means that a vortex located in this region will be pulled 
inside the sample, it is possible to allocate vortices in this
region and there is also a region iii) near the sample surface that
repels vortex entrance, $G$ increases for increasing $x$. The 
existence of regions ii) and iii) means a reinforcement of the 
surface barrier induced by vortex penetration and allows magnetic 
field intervals of metastable states. This energy barrier 
reinforcement due to vortex entrance is more important in 
mesoscopic superconductors because, in small samples, vortices are 
confined by the potential well generated by the sample surface and 
even a vortex fixed in the center of the sample is very close to the 
surface. For example in an 
$10 \lambda \times 10 \lambda$ square sample, vortices are 
constrained to be located around the center of the sample at
$x=5 \lambda$, because of its interaction with the surface currents. 
As a consequence vortices stay at $x \approx 5 \lambda$ 
generating a new surface barrier and it is necessary an
important magnetic field increase to allow new vortex
penetration. For macroscopic samples the situation 
is quite different, there is a nearly continuous vortex penetration. 
In macroscopic samples the vortices that are inside the sample are
not  confined and they do not have serious restrictions in their 
movement since they can be allocated very far from the surface. 
In this case, a small increase of the magnetic field is enough to accommodate 
new vortex lines, as can be observed in Figure 5(a).

From the analysis of Figure 9 and the previous 
discussion we can conclude that in mesoscopic samples 
there are preferred values of magnetic field 
for vortex penetration, in this case the process of vortex
entrance is discontinuous in contrast with the continuous macroscopic
regime. This behavior is a consequence of the barrier to vortex 
entrance that appears after each penetration event. In this way 
we can define a second penetration field $H_{p2}$, a third 
penetration field $H_{p3}$ and so on.  We observe that after 
increasing the size of the sample the vortex penetration becomes 
continuous ($H_{p3} \rightarrow H_{p2} \rightarrow H_p$).

The exact delimitation of the macroscopic and mesoscopic regimes 
depends on the geometry of the sample. We will now study in detail 
the behavior of a thin film because is a simpler case with only 
one significant length scale. The size dependence of the penetration 
fields $H_p, H_{p2}, H_{p3}$ obtained numerically from the TDGL equations
in a thin film using the S-I boundary condition are
summarized in Figure 11(a).  For $d > 15 \lambda$ a continuous 
vortex penetration is observed, this is the region of a 
macroscopic behavior. The mesoscopic region is located at 
$ 2 \lambda \lessapprox d \lessapprox 15 \lambda$ in which
several penetration fields $H_{p_i}$ can be distinguished. For 
$d < 2 \lambda$ there is a gradual transition to the normal state
without vortices. 
Figure 11(b) shows the sample size 
dependence of the first, second and third penetration field of 
thin films using an S-N boundary condition. For the S-N 
boundary the scales are shifted to larger sizes 
since the shielding length is larger ($\delta_{SN}>\delta_{SI}$), as
we discussed before.
The macroscopic behavior appears at $d > 18 \lambda$, the 
mesoscopic region is located between $ 5 \lambda \lessapprox d
\lessapprox 18 \lambda$ and 
for $d < 5 \lambda$ a continuous transition to the normal state appears.

It is interesting to note that $H_p$ is size dependent in the 
mesoscopic region. 
This size dependence can be explained by considering in Equation (13)
the effect of two surfaces that are separated by a small distance $d$. 
The term generated by the magnetic field 
$H \exp(-x/ \lambda)$ now becomes $H\cosh[(x-d)/ \lambda]/\cosh(d/
\lambda)$ due to the proximity of the other surface, the
magnetic force that pulls the vortex inside the sample decreases 
when the film thickness ($d$) decreases. The
image term also changes because the vortex lines that are trying to
enter the sample are now near both surfaces and new image
lines are necessary in order to satisfy the boundary condition at
both surfaces. The application of the image method to parallel 
surfaces gives an infinite number of images, but the net effect 
is the appearance of an attractive interaction to the new surface. 
Then, in the mesoscopic region, there is also a decrease in the 
net attractive image force. 
If we consider only three relevant terms in the image forces, the 
normalized force $f=(4 \pi / \Phi_o H_{c2}) \lambda (\delta G/ \delta
x)$ that feels a vortex line that is trying to enter when the
sample is in the Meissner state can be estimated by:
\begin{eqnarray}
f= H_a \frac{\sinh(\frac{x-d/2}{ \lambda})}{\cosh(d/2 \lambda)}+
\frac{1}{\kappa^2} 
 K_1(\frac{2x}{\lambda})- \nonumber \\
 \frac{1}{\kappa^2} K_1(\frac{2d-2x}{\lambda})+
 \frac{1}{\kappa^2} K_1(\frac{2d+2x}{\lambda}) 
\end{eqnarray}
where $d$ is the thickness of the film and we have chosen $f$ positive
when it repels the vortex entrance. As we analyzed before, $H_p$
is usually obtained from $f|_{x= \xi}=0$.
\begin{figure}[htb] 
\centerline{\epsfxsize=8cm \epsfbox{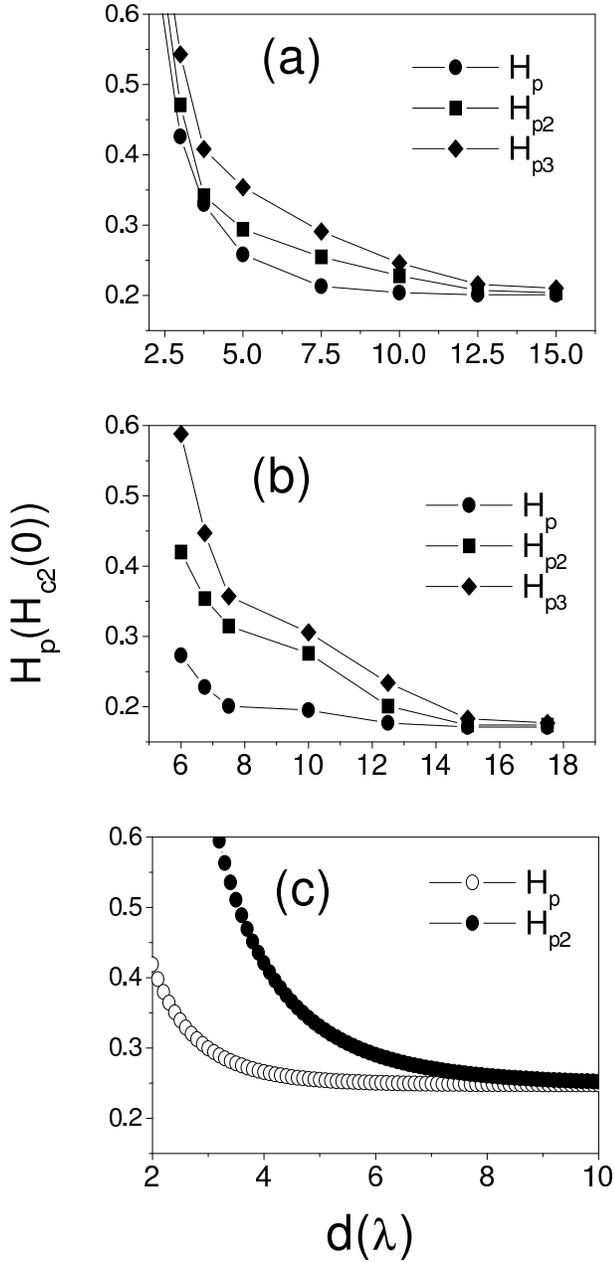}}
\caption{Penetration fields obtained from the magnetic
curves shown in Figures 11 and 12. We plot the first, second and
third penetration fields at different film thickness for (a) S-I
and (b) the S-N boundary condition. 
$H_{p3} \neq H_{p2} \neq H_p$ is typical of a mesoscopic behavior.
For large film thickness, in the region of the macroscopic
behavior, a continuos entrance of vortices is recovered 
($H_{p3} \rightarrow H_{p2} \rightarrow H_p$). Figure (c) shows an
estimation of $H_p$ and $H_{p2}$ using the image method and
considering the influence of the sample size in the field profile.} 
\label{fig11} 
\end{figure}\noindent
This condition leads to the following expression for $H_p$: 
\begin{eqnarray}
H_p =\frac{[K_1(\frac{2 \xi}{\lambda})-K_1(\frac{2d-2 \xi}{\lambda})+
  K_1(\frac{2d+2 \xi}{\lambda})]\cosh(\frac{d}{2 \lambda})}{\kappa^2
  \sinh(\frac{d-2 \xi}{2 \lambda})}
\end{eqnarray}

We have evaluated this behavior of $H_p (d)$ in Figure 11(c). 
As it can be observed, when we decrease the thickness of the film the
 repulsive magnetic force decreases faster than the attractive term,
 and the $H_p$ value increases. 
In Figure 11(c) we also estimate the behavior of the second
penetration field $H_{p2}$ as a
function of the sample size. The approximate expression used was
obtained including in the Equation (15) the extra terms related
with the presence of a vortex line inside the sample as we did in 
Equation (14). For simplicity we have considered one vortex 
located at the middle of the sample at $x=d/2$. Under this 
condition $H_{p2}$ becomes:
\begin{eqnarray}
H_{p2}&=&[K_1(\frac{2 \xi}{\lambda})-K_1(\frac{2d-2 \xi}{\lambda})+
  K_1(\frac{2d+2 \xi}{\lambda})+ \nonumber \\
  &+& K_1(\frac{d- 2 \xi}{2 \lambda}) + K_1(\frac{d+ 2 \xi}{2
  \lambda}) - \nonumber \\
  &-& K_1(\frac{3d+ 2 \xi}{2 \lambda})]\frac{\cosh(\frac{d}{2
  \lambda})}{\kappa^2
  \sinh(\frac{d-2 \xi}{2 \lambda})}
\end{eqnarray}
this expression also reproduces the $H_{p2}$ sample size dependence 
observed in the numerical simulations.

\section{SUMMARY}

We have presented results on the study of the
magnetization curves and the surface barrier for type 
I and type II superconductors.
Our results show that the strength
of the surface barrier depends on the boundary. 
 If the interface is of the S-I 
type, the vortices that enter find a higher barrier for
penetration since
the order parameter at the surface has to go through an intermediate
state of $\Psi=0$, while for the S-N 
boundary condition the vortex entrance occurs more smoothly since
the $\Psi=0$ condition is already fulfilled by the interface.
In this later case the surface barrier is only due to the Meissner shielding
currents and the penetration field $H_p$ agrees well with the
Bean-Livingston value, while in the S-I case $H_p$ is higher and
dependent on $\kappa$. Superconductors with more realistic boundary
conditions should lie in between these two cases.

We also characterized the reinforcement of the surface barrier due to the 
presence of vortex lines inside the sample in mesoscopic superconductors. 
We show that these new barriers allow for the existence of 
metastable states of constant vorticity as a function
of magnetic field. Each metastable state becomes unstable
at the $i$-th penetration field $H_{pi}$ in which one vortex enters 
in each side of the sample and the magnetization has a discontinuous
jump. We study the 
magnetization curves at different sample dimensions 
and we obtain the sample size 
dependence of the first, second and third penetration fields. 
We finally show that for
sufficiently large sample sizes the continuous macroscopic 
regimen is recovered, i.e. 
$H_{p3} \rightarrow H_{p2} \rightarrow H_p$.

\acknowledgements

We acknowledge helpful discussions with Arturo L\'{o}pez and 
Niels Gr\o nbech-Jensen. 
A.D.H. acknowledges Oscar Ar\'{e}s and C. Hart for useful
comments and help and the Centro Latino-Americano de F\'{\i}sica 
(CLAF) for financial support.  D.D. acknowledges
support from Conicet and CNEA (Argentina). We also acknowledge
the financial support for this project from 
ANPCyT and Fundaci\'{o}n Antorchas.


\begin{references}
\bibitem{nota}
Only out of equilibrium we find that  structures of
coexisting normal and superconducting phase can be nucleated, 
as was found in Ref.~\onlinecite{frah,liu}.
\end{references}
\end{document}